# Monte Carlo Simulation of Carrier Diffusion in Organic Thin Films with Morphological Inhomogeneity

S. Raj Mohan[1*], Manoranjan P. Singh[2], M. P. Joshi[1], L. M. Kukreja[1]

[1]Laser Materials Processing Division, [2]Laser Plasma Division,

Raja Ramanna Centre for Advanced Technology, Indore, India, 452013.

[*] Corresponding author.  Tel.:  +91 7312488361;  fax: +91 7312488300

E-mail address:  raj@rrcat.gov.in; rajmho@gmail.com

**Abstract**

Monte Carlo simulation was carried out to understand the influence of morphological inhomogeneity on carrier diffusion in organic thin films. The morphological inhomogeneity was considered in the simulation by incorporating the regions of low energetic disorder in a host lattice of high energetic disorder which decreases the overall energetic disorder of the system. For the homogeneous films, the carrier diffusion was found to decrease upon decreasing the energetic disorder. In contrast to this, in the case of inhomogeneous films the carrier diffusion enhanced upon decreasing the overall energetic disorder, up to an optimum value and beyond which the carrier diffusion decreased. Through our simulation, we observed that the behavior of carrier diffusion in the inhomogeneous case is due to the morphology dependent carrier spreading, which acts in addition to the thermal and non-thermal field assisted diffusion mechanisms. This morphological dependence of carrier spreading arises due to the generation of packets of carriers with different jump rates, which is after effect of slow relaxation of the carriers generated in the less disordered regions of inhomogeneous system. Our simulation of morphology dependent carrier spreading and its influence on the basic diffusion process provide deeper insight into the charge transport mechanisms in organic thin films.

0

**Introduction**

Disordered organic materials have attracted researchers from various disciplines due to the potential ability of these materials in developing various cost effective organic optoelectronic devices, like organic solar cells, organic light emitting diodes etc[1]. The presence of high disorder along with weak intermolecular force of attraction makes disordered organic materials stand apart from the classical semiconductors[1-3]. Compared to inorganic semiconducting materials, the materials in this class almost retain their independent identity in the bulk form. Hence, instead of bands, valance and conduction bands as in the inorganic counterpart, these materials in the bulk have localized states, that are subjected to positional and energetic disorder[1-4]. In light of these remarkable differences, between the disordered organic materials and inorganic semiconductors, the use of physical models developed for inorganic semiconductors may not be suitable for explaining the observed phenomenon in disordered organic materials[1-4]. For example, compared to inorganic semiconducting materials the charge transport in disordered organic materials occurs by means of hopping among localized states. Hence, the magnitude of mobility is very low and shows a Poole-Frenkel and non-Arrhenius type electric field and temperature dependence, respectively[4]. Diffusion of the carriers is another phenomenon where the disordered organic materials deviate from the conventional Einstein's law relating the mobility and diffusion coefficient[3,5-9]. Significant deviation from Einstein's law is observed upon increasing the disorder and electric field[5-7]. The deviation has been attributed to the enhanced diffusion due to the non-thermal field assisted diffusion, arising due to the wide difference in jump rates of carriers occupying the top and bottom of density of states (DOS)[6,10,11]. The diffusion of carriers has significant influence on the optoelectronic properties of disordered organic materials[1-9] and hence a better understanding on the process of carrier diffusion is crucial for the design of efficient organic optoelectronic devices[12-15]. Recent theoretical and experimental reports[16-21] on carrier diffusion in disordered organic materials assert that the valuable knowledge on the carrier diffusion that has been gained through decades of research needs more refinement to deal with various types of samples. Moreover, the advent of new class of materials[22-24] and many new methodologies[25-28] for improving the charge transport properties also demand further investigation on carrier diffusion[21]. Investigation on carrier diffusion is very important for not only to achieve deeper understanding on the device physics but also for developing theoretical frameworks that can account charge transport in finer details.

The influence of film morphology on optical and electrical properties has been investigated extensively as this understanding is indispensable for the design and development of efficient organic optoelectronic devices[25-32]. One of the prime motives behind tailoring the film morphology is to incorporate more structural order in an otherwise highly disordered homogeneous system for providing percolation pathways and thereby enhancing the charge carrier mobility[25-28]. These morphologically tailored films contain a blend of ordered regions and disordered regions, where the energetic disorder is low and high respectively[25-32]. Hence, these morphologically tailored films cannot be treated as homogeneous rather they are inhomogeneous[25-32]. The charge transport in such system occurs through a mixture of ordered and less ordered regions. In such cases, the energetic disorder seen by the carrier changes intermittently and the influence of energetic disorder on charge transport become complex[25-32]. In this context, it is pertinent to ask as how is diffusion of carriers affected by the presence of inhomogeneity in the sample. In particular, does the morphology of the sample give rise to any additional mechanism of diffusion. In order to answer these we study the diffusion of carriers in morphologically different disordered system using Monte Carlo simulation[4]. This study essentially investigates the influence of spatial fluctuation in energetic disorder on carrier diffusion. The influence of morphology on diffusion of carriers is investigated by studying the temporal evolution of diffusion coefficient along the applied field direction as a function of concentrations of ordered regions of low disorder (COR) embedded inside a highly disordered host lattice. In this study, the influence of morphology on diffusion is inferred from the transient diffusion data. Earlier reports[30-32] suggest that the effective energetic disorder seen by the carrier decreases with the increase in COR. Compared to the homogeneous medium, the temporal evolution of diffusion coefficient calculated along the applied field direction for the inhomogeneous medium shows unexpected behavior upon decreasing the overall energetic disorder seen by the carrier. The unexpected behavior of diffusion coefficient is explained using a morphology dependent carrier spreading mechanism that acts on carrier packet, in addition to the usual thermal and non-thermal field assisted diffusion mechanisms. The origin of this morphology dependent carrier spreading mechanism is due to the influence of film morphology on carrier relaxation process. Finally, the ramifications of the morphology dependent carrier spreading mechanism are also discussed.

**Details of simulation**

A 3D array with size 70x70x10000 along *x*, *y* and *z* direction is considered as the lattice. The size of the lattice is judged on the basis of our intention to change the lattice morphology and also by taking into account the available computational resources. *Z* direction is taken as the direction of the applied field. The lattice constant *a* = 6Å is taken for the whole set of simulation[4,30-32]. The site energies are assumed to be correlated with Gaussian distribution[31,32]. Simulation is performed on an energetically disordered lattice with the assumption that the hopping among the lattice sites is governed by Miller-Abrahams equation[4,30-32] in which the jump rate ($\upsilon_{ij}$) of the charge carrier from the site *i* to site *j* is given by

$$\upsilon_{ij} = \upsilon_o \exp\left[-2\gamma a \frac{\Delta R_{ij}}{a}\right] \begin{bmatrix} \exp\left[-\dfrac{\varepsilon_j' - \varepsilon_i'}{kT}\right] &, & \varepsilon_j' > \varepsilon_i' \\ 1 &, & \varepsilon_i' > \varepsilon_j' \end{bmatrix} \quad (1)$$

where $\Delta R_{ij} = |R_i - R_j|$ is the distance between sites *i* and *j*, $\varepsilon_i'$ and $\varepsilon_j'$ are the effective energies of the site *i* and *j* which include the electrostatic energy, *a* is the intersite distance, *k* is the Boltzmann constant, *T* is the temperature in Kelvin and $2\gamma a$ is the wave function overlap parameter which governs the electronic exchange interaction between sites. Throughout the simulation the positional disorder is neglected with the value of overlap parameter[4,30-32] taken to be $2\gamma a=10$. This will be referred as homogeneous lattice (HL) hereafter. In this study, the Monte Carlo simulation is based on single carrier approach which is appropriate for very low carrier concentration where the influence of space charge effects[33,34] can be neglected. The charge carrier is injected randomly on to the first plane of the lattice, which is then allowed to hop in the presence of applied electric field. Every injected carrier is allowed to hop until it covers a sample length of *6μm*. Simulation is performed by averaging over ten thousand carriers with one lattice realization per carrier. Position of the carrier as a function of time is monitored and the temporal evolution of diffusion coefficient is calculated by using the equation,

$$D_z = \frac{\left\langle (z - \langle z \rangle)^2 \right\rangle}{2t} \quad (2)$$

Diffusion coefficient presented in the manuscript is calculated along the applied field direction ($D_z$ calculated along $z$ direction) at a temperature T=300K. According to the report by Nenashev *et al*[16] site energies around $-2\sigma^2/kT$ are determinant for field dependent diffusion and hence very important. The report suggests that to have the required site energies (~-6σ for T=300K) in the simulation a lattice of bigger size is essential. To handle such big arrays huge computational resources are required. Our investigation (data not shown) suggests that accurate value of diffusion coefficient can also be calculated using smaller lattice (for eg. 70x70x10000, lattice size use in this work) provided averaging should be carried over large number of lattice realizations. The averaging over large number of lattice realizations will bring the influence of deep site energies and hence provides accurate values of diffusion coefficient using smaller lattice sizes. Our investigation has shown that the averaging carried over ten thousand carriers with one lattice realization per carrier used in this manuscript brings in the influence of required deep site energies[16] and hence the proper estimation of diffusion coefficient.

A simplified model is adopted to study the influence of morphology on the carrier diffusion. For inhomogeneous system the lattice morphology is varied by embedding cuboids of ordered regions randomly inside a highly disordered host lattice[30-32]. The size of ordered regions was also chosen randomly. Size of ordered regions is limited to a maximum size of 25x25x40 sites along *x*, *y*, and *z* directions, if not mentioned otherwise. The size of ordered region is varied by changing the maximum size of cuboid which is inserted to obtain the required COR. Size of cuboids is chosen such that a nanoscale morphology[25-32] can be obtained and also the morphology can be varied upon changing the COR. Energetic disorder inside the ordered region is kept low compared to the host lattice. Such a highly disordered lattice with embedded ordered regions is justified because the organic/polymer films employed in practical devices are mostly morphologically tailored (either intentionally or with aging) and therefore contains regions of low and high disorder[30-32]. The site energies inside the host lattice and ordered regions are assumed to be correlated and follow Gaussian distribution. Throughout the simulation, same mean energy of the DOS, 5.1eV, is assumed for the DOS of host lattice and ordered regions. Energetic disorder for the host lattice is taken to be 75meV, (a typical value of energetic disorder seen in the homogeneous disordered organic materials), and for the ordered regions is taken to be 15meV (five times less compared to that for the host lattice). Earlier reports have even suggested a ten-fold reduction of energetic disorder inside the polycrystalline regions[35]. The above lattice will be referred as inhomogeneous lattice (IHL) henceforth. As a special case of IHL (*Mixed distribution case* (Mixed Dist.)), simulation is also performed on a host lattice where the site energies of host lattice were randomly chosen from a narrow ($\sigma = 15meV$) and broad Gaussian distribution ($\sigma = 75meV$). In this case, there is no regions of low disorder instead there will be sites of low disorder spread randomly, with weight governed by COR, inside the highly disordered host lattice.

We assume hopping transport, governed by Miller-Abhrams equation, inside the ordered regions. In some of the earlier reports[36,37], authors propose band transport inside the ordered regions/aggregates. However, the band transport inside the low disordered regions /aggregates has not been established for the existing wide variety of organic materials. Moreover, the presence of energetic disorder inside the aggregates[35] along with the weak intermolecular force supports hopping transport than a band transport. Hence, we assume hopping transport inside the ordered region too and we believe this to be more appropriate. Carrier encountering the interface between the disordered region and ordered region faces only the fluctuation in the energetic disorder between these two regions. Practically, the interfacial properties at organic–organic interface are very complicated[38] and have significant influence on the optoelectronic properties. The nature of the organic-organic interface are characterized by several parameters such as degree of crystallanity, polymorphism, roughness and dimension of the interfacial regions and presence of impurities. These parameters in turn depend on several other factors such as chemical nature of the materials, techniques adopted for the growth of interface and growth parameters such as temperature, solvent, spin speed, concentration and annealing time/temperature[38]. Thus, incorporating these complex interfacial properties accurately in simulation is very difficult. In this study, a simplified organic–organic interface[33,34,39,40] is considered for investigating the influence of film morphology on carrier diffusion. Moreover, with this simplified approach the simulation exclusively considers the influence of spatial fluctuation in energetic disorder, which is certainly present in the morphologically tailored samples and have significant influence on the charge transport. Like in the homogeneous lattice, every carrier is injected randomly on to the first plane of the lattice, which is then allowed to move in the presence of applied electric field. We assume uniform carrier generation in both ordered and disordered region. This is an idealized behavior, not practical in real materials and is adopted to bring out the exclusive influence of spatial fluctuation of energetic disorder on charge transport[40]. As explained above, the temporal evolution of diffusion coefficient along the applied field direction is simulated by varying COR, size distribution of the ordered region and the applied electric field.

**Results and Discussions**

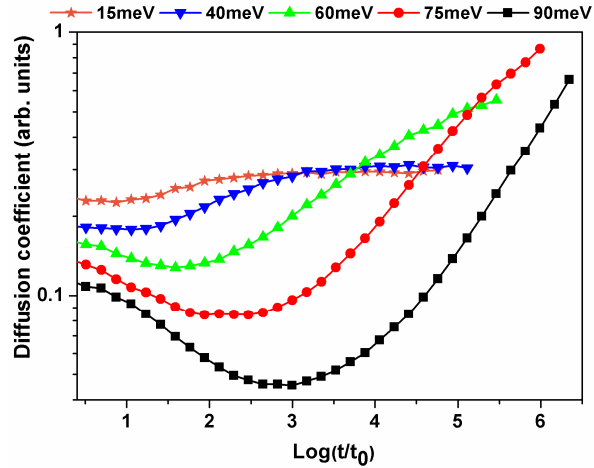

Figure 1. Temporal evolution of diffusion coefficient parametric with various value of energetic disorder for HL. All simulation were carried out at E=6.4x10$^5$V/cm and T=300K.

Simulations were carried out to obtain the temporal evolution of diffusion coefficient in both HL and IHL. In case of IHL and for most of the cases of HL the diffusion coefficient has not attained the steady state value even at long time ($Log(t/t_0)$~6). In what follows we essentially discuss the transient behavior of diffusion coefficient for the respective cases. We will briefly comment about the steady state diffusion whenever possible. In order to emphasize the observations on IHL, we first explain the temporal evolution of diffusion obtained for HL. This is shown in figure 1 for various values of energetic disorder *(time normalized with the dwell time of a lattice without disorder, $t_0$)*. For all the values of energetic disorder, the calculated value of diffusion coefficient decreases initially and reaches a minimum value (valley point) before it starts increasing with time[6]. After reaching the valley point, the rate at which the diffusion coefficient increases with time also decreases with decrease in the energetic disorder. Time taken to reach the steady state increases with increase in the energetic disorder. For example, the temporal evolution of diffusion coefficient for 90meV has not shown even sign of reaching the steady state behavior even at long time ($Log(t/t_0)$~6). It is well known that the diffusion coefficient contains contributions from the thermal and non-thermal field assisted diffusion[6,10]. The non-thermal field assisted diffusion is field and energetic disorder dependent and arises due to the wide difference in the jump rates of the carriers located at the bottom and top of the DOS[6,10]. Consequently, higher the energetic disorder larger is the contribution from the non-thermal field assisted diffusion. This is because of the large difference in jump rates of carriers located at the bottom and top of DOS at higher values of energetic disorder. After attaining the valley point, the observed higher rate of increase of diffusion coefficient (faster spreading of carrier packet) with time for higher values of energetic disorder can be attributed to higher contribution from the non-thermal field assisted diffusion[20]. It should be mentioned here that the aforesaid rate of increase in diffusion pertains to the time domain where the carriers have not approached the steady state. As the steady state is approached, the rate of increase in diffusion coefficient tends to zero, as observed for lower values of energetic disorder. After explaining diffusion in HL we look at the same in IHL.

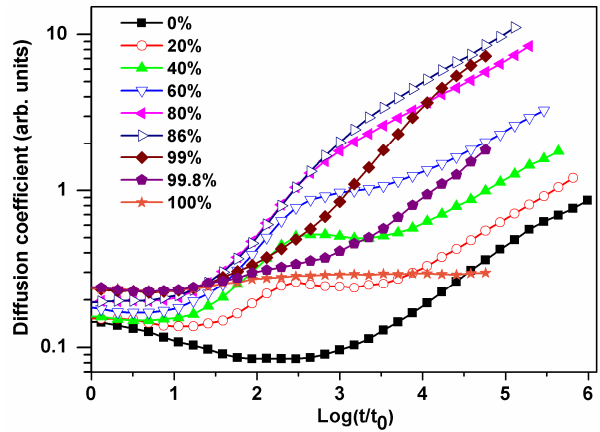

Figure 2. Temporal evolution of diffusion coefficient parametric with COR for IHL. All simulation were carried out at $E=6.4 \times 10^5 V/cm$ and $T=300K$ with maximum size of ordered region is 25x25x40 along x, y and z directions.

In IHL, the morphology of the lattice is modified by incorporating ordered regions inside the host lattice. It is known from our earlier reports[30-32] that the overall energetic disorder seen by the carrier decreases with increase in COR. Hence, the temporal evolution of diffusion upon increasing COR is expected to behave in a similar manner as observed in HL upon decreasing the energetic disorder. Figure 2 shows the temporal evolution of diffusion coefficient for various COR. For comparison, the temporal evolution of diffusion coefficient for a pure host lattice ($\sigma$=75meV) is also shown. In each case, the diffusion coefficient initially decreases and reaches a minimum value (valley point) before it starts increasing with time. After reaching the valley point the diffusion coefficient in all the cases increases with time but shows an intermediate regime at low COR. In the intermediate regime, the rate of increase in diffusion coefficient with time is small. The intermediate regime gradually vanishes upon increasing COR. After the intermediate regime, the diffusion coefficient increases further with time but at a slower rate. This rate is approximately same as that of pure host lattice. At higher COR, even though the intermediate regime vanishes, the rate of increase in diffusion coefficient at longer times is smaller compared to the rate attained immediately after the valley point. Immediately after the valley point, the rate of increase in diffusion coefficient increases with increase in COR, attaining a maximum at ~85% COR. As COR is increased beyond 85% a decrease in the rate of increase of diffusion coefficient is observed. Further more at higher COR, the diffusion coefficient does not show steady state behavior, as observed in HL with low energetic disorder. This shows that the dynamic equilibrium is delayed upon incorporating ordered regions inside the host lattice and hence suggests a possible influence of film morphology on dynamic equilibrium. As compared to HL the behavior of the diffusion coefficient in IHL is quiet intriguing *(mechanism explained later)*.

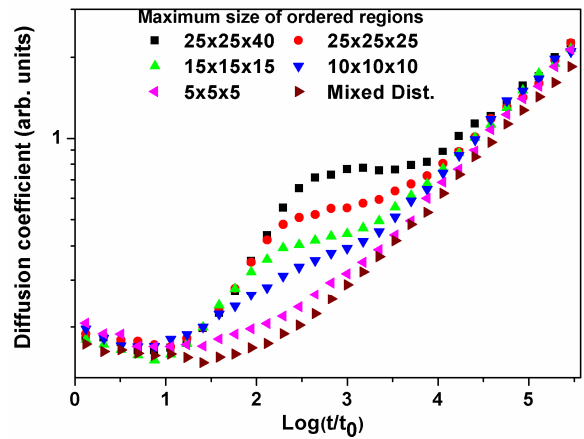

Figure 3. Temporal evolution of diffusion coefficient upon decreasing the maximum size of ordered region for a COR of 50%. All simulation were carried out at $E=6.4 \times 10^5 V/cm$ and $T=300K$.

The diffusion coefficient increases with increase in COR, i.e. diffusion coefficient increases with the decrease in the overall energetic disorder seen by the carrier whereas in HL results would suggest that this should be just opposite where diffusion coefficient decreases with decreases in energetic disorder. This makes one infer that in IHL the carrier diffusion is governed by a morphology dependent factor, which acts on the carrier packet in addition to the contribution from the thermal and non-thermal field assisted diffusion[6,7,10,11]. In order to establish the role of a morphology dependent factor in governing the carrier diffusion in IHL, the temporal evolution of diffusion coefficient is simulated by varying the maximum size of the cuboid which is inserted to achieve a fixed COR, thereby generating different lattice morphologies for fixed COR. Figure 3 shows how the observed features of temporal evolution of diffusion coefficient of an IHL changes upon decreasing the size of ordered region for a COR of 50%. Upon decreasing the maximum size of ordered region, the rate of increase of diffusion immediately after the valley point decreases while at the same time intermediate regime also vanishes. This clearly establishes the fact that the carrier diffusion in IHL is controlled by a morphology dependent factor.

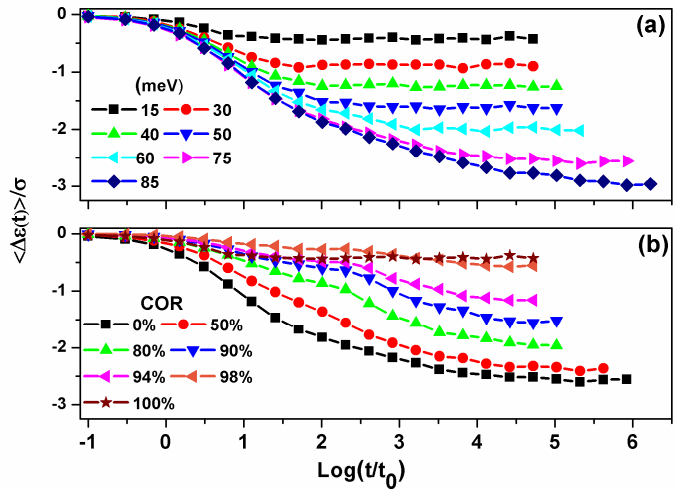

Figure 4. Temporal evolution of mean energy of carriers measured with respect to the center of DOS (a) HL, parametric with various values of energetic disorder (b) IHL, parametric with COR. Maximum size of ordered region is 25x25x40 along *x*, *y* and *z* directions. Time normalized with the dwell time of a lattice without disorder, $t_0$. All simulation were carried out at E=6.4x$10^5$V/cm and T=300K, with carriers started at 5.1eV.

In order to understand the origin of morphology dependent carrier diffusion we first analyze the relaxation of carriers for both HL and IHL. Carriers generated in a Gaussian DOS will undergo relaxation to the bottom of DOS and the mean energy of the carrier packet saturates at long times[3,4] to a constant value ($<\varepsilon_\infty>$). The value of $<\varepsilon_\infty>$ depends on energetic disorder, applied electric field and temperature. At zero electric field the equilibration energy[3,4] measured from the mean of DOS is given as $<\varepsilon_\infty> = -\frac{\sigma^2}{kT}$. In the presence of finite electric field $<\varepsilon_\infty>$ is higher. The time ($\tau_{rel}$) at which carrier packet attains the equilibration energy also strongly depends on energetic disorder and temperature[3,4] ($\tau_{rel} \propto exp\left(\frac{B\sigma}{kT}\right)^2$). Figure 4(a) shows the temporal evolution of the mean energy of carrier packet for HL, parametric with energetic disorder. The data shown in Figure 4(a) is obtained for constant electric field strength (E=6.4x10$^5$V/cm) and temperature (T=300K) with carriers generated at 5.1eV. For all values of energetic disorder carrier relaxes smoothly towards the bottom of DOS and at long time the mean energy of the carrier attains the respective steady state value, $<\varepsilon_\infty>$. From Figure 4a it is clear that $<\varepsilon_\infty>$ increases upon decreasing the value of energetic disorder while $\tau_{rel}$ also decreases concomitantly. This confirms that upon decreasing the overall energetic disorder the carriers quickly relax to attain the dynamic equilibrium and the mean energy of carrier packet climbs towards the mean of DOS. In IHL, carriers are generated inside the host lattice as well as in the ordered regions. Carriers generated in host and ordered regions are expected to relax to the bottom of DOS, which is present mostly in the disordered host lattice. Figure 4(b) shows the temporal evolution of the mean energy of carrier packet with time for IHL, parametric with COR. For comparison, the temporal evolution of mean energy of the carriers for pure host lattices (HL with energetic disorder of 75meV and 15meV, special cases of COR equal to 0% and 100% respectively) is also shown. Upon increasing COR the temporal evolution of the mean energy of carrier packet shows a shoulder before it saturates to $<\varepsilon_\infty>$ at long time. The presence of shoulder becomes prominent with increase in COR. The value of $<\varepsilon_\infty>$ increases with increases in COR while the value of $\tau_{rel}$ does not change much with COR. The presence of the shoulder and its prominence at higher COR, suggest that in IHL the mean energy of the carriers decreases slowly with time as compared to HL. This implies that the relaxation of the carrier packet in IHL becomes slower with increase in COR. As the inclusion of ordered region decreases the overall energetic disorder of the system, one would expect a faster relaxation of carriers to $\varepsilon_\infty$ with the increase in COR.

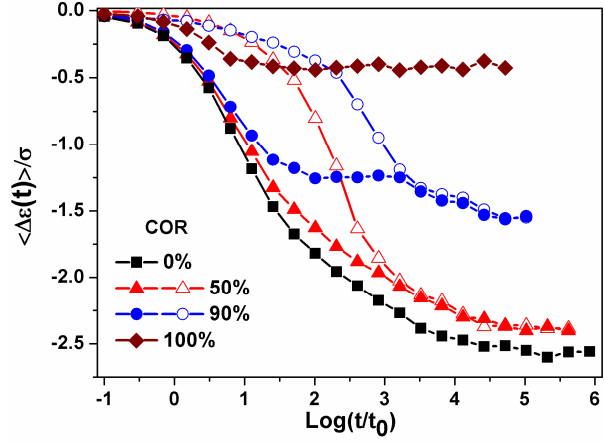

Figure 5. Temporal evolution of mean energy of carriers, generated specifically inside highly disordered (▲,●) and ordered regions of IHL (△,○). Maximum size of ordered region is 25x25x40 along x, y and z directions. Temporal evolution of mean energy of carriers for HL with $\sigma$=75meV (■) and $\sigma$= 15meV (♦) are also shown for comparison. Time normalized with the dwell time of a lattice without disorder, $t_0$. All simulation were carried out at E=6.4x10$^5$V/cm and T=300K, with carriers started at 5.1eV.

In order to understand the origin of the shoulder and hence the slower relaxation of carriers observed in IHL we have investigated the relaxation of the carrier generated specifically inside the highly disordered regions and ordered regions of the IHL. Figure 5 shows the temporal evolution of the mean energy of carrier packet generated in the regions of high disorder (host lattice) and in the regions of less disorder (ordered regions). For comparison, the evolution of the mean energy of the carriers inside a pure host lattices (HL with energetic disorder of 75meV and 15meV, special cases of COR equal to 0% and 100% respectively) is also shown. It is observed that in the beginning the mean energy of the carriers generated inside the ordered region relax slowly compared to the carriers generated inside the highly disordered region. As COR is increased, the relaxation of mean energy of the carrier generated inside the ordered regions becomes further slower. The relaxation of the carriers generated inside the ordered region, even at high COR, is clearly different from HL with $\sigma$=15meV. At a sufficiently long time ($Log(t/t_0)$~>3) both curves merge to form a single curve, suggesting that the relaxation of carriers generated in ordered and highly disordered regions happens in a similar fashion. Thus, it is clear that carriers generated in the ordered region have a slower pathway of energy relaxation. This results in the formation of shoulder (slower relaxation of carriers) in Figure 4(b) for IHL, in which carriers are generated inside the highly disordered and ordered regions. Figure 6 clearly shows that the difference in energy relaxation of charge carriers generated in different regions is sensitive to the size of the ordered region – bigger the size of ordered region more prominent is the difference. This suggests that the difference in relaxation of the carriers generated inside the ordered region and highly disordered regions is entirely dependent on the film morphology.

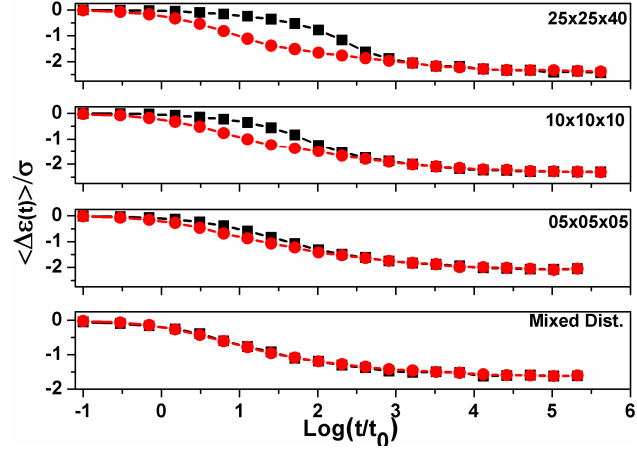

Figure 6. Temporal evolution of mean energy of carriers, generated specifically inside highly disordered (●) and ordered regions of IHL (■), upon decreasing the maximum size of ordered regions. All simulation were carried out at E=6.4x10$^5$V/cm and T=300K, with carriers started at 5.1eV and COR is 50%.

In order to understand the origin of slow relaxation of carriers generated inside the ordered region we followed the path of such carriers. Figure 7(a) shows the temporal evolution of fraction of low disordered sites and high disordered sites visited by the carriers generated inside the ordered region for a COR =50%. Initially, the fraction of low disordered sites visited by the carrier decreases very slowly. Remarkable decrease in the fraction of low disordered sites visited by the carriers is observed when $Log(t/t_0)>2$ with a concomitant increase in the fraction of high disordered sites visited by the carrier. At long times, as the carriers relax to the bottom of DOS, the fraction of high disordered sites visited by the carriers become higher than fraction of low disordered sites visited by the carriers. As the carrier generated inside the ordered regions is around 5.1eV and initially as they remain in the ordered regions for long the slower energy relaxation of such carriers is clearly justified. The remarkable decrease in average energy of carriers generated in the ordered region (for ordered region size 25x25x40, shown in figure 5) is observed at the same time when a remarkable decrease in the fraction of low disordered sites visited by the carriers is observed, i.e. when $Log(t/t_0)>2$. Figure 7(b) shows the variation in the temporal evolution of fraction of low/high disordered sites visited by the carriers upon decreasing the size of ordered region. Remarkable decrease/increase in the fraction of low/high disordered sites visited by the carriers happens at shorter times ($Log(t/t_0)>0$) when the size of ordered region is decreased. This clearly establishes the role of morphology in energy relaxation of carriers. In concise, as the size of ordered regions is large, film morphology allows the carriers to remain in the ordered region for long and hence carriers scan the energy space less effectively. This results in the slower relaxation of the carriers. When the size of ordered region is reduced the carrier generated in the ordered region could scan the energy space more efficiently which is supported by the fast reduction in the fraction of low disordered sites visited by the carriers and hence faster relaxation of the carriers. This explains why the difference in energy relaxation between the carriers generated in the ordered and highly disordered regions vanishes upon decreasing the size of ordered region.

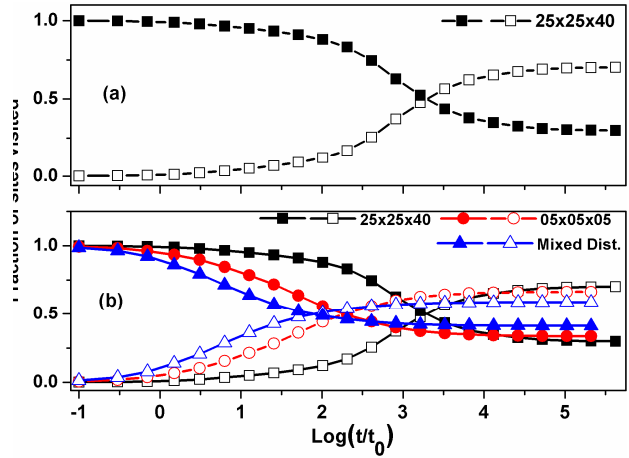

Figure 7. Temporal evolution of fraction of low disordered sites (closed symbols) and high disordered sites (open symbols) visited by the carriers generated inside the ordered region (a) For ordered region of maximum size 25x25x40 (b)Comparison upon decreasing the maximum size of ordered regions. All simulation were carried out at E=6.4x10$^5$V/cm, T=300K and COR is 50%.

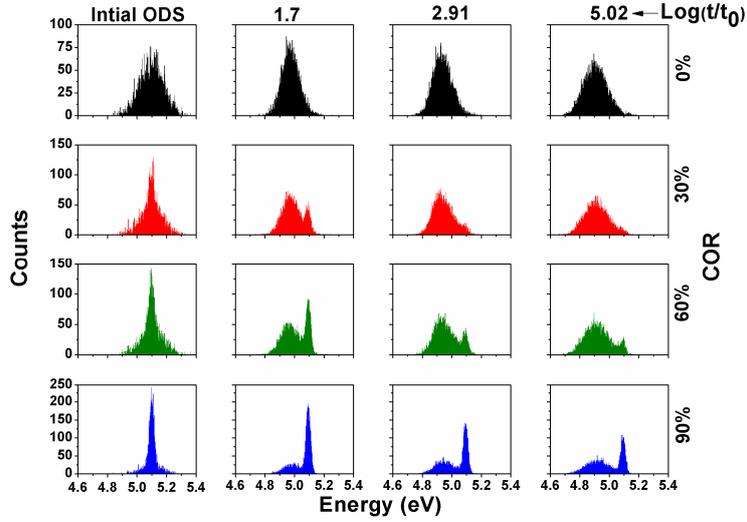

Figure 8. Histogram representing the temporal evolution of the occupation of DOS for a IHL at various COR. All simulation were carried out at E= $6.4 \times 10^5$ V/cm and T=300K.

Further investigation is carried out to understand how the slower relaxation of the charge carriers in the IHL influences the temporal evolution of diffusion coefficient upon decreasing the overall energetic disorder seen by the carrier. Figure 8 shows the density of occupied states (ODS), captured at different times (time normalized with $t_0$), for the carriers generated in the IHL for various COR. In case of pure host lattice (0% COR), the carrier packet relaxes smoothly to the lower energy side. When the ordered regions are embedded into the host lattice the carriers are generated both inside the highly disordered (host lattice) and ordered regions. Since the energetic disorder inside the ordered region is very low ($\sigma$~15meV) the ODS of carriers generated inside the ordered region is narrow around 5.1eV, mean energy of DOS. Hence, initially the ODS for IHL has a sharp peak around 5.1eV. Upon increasing COR the initial peak becomes more sharper. As time elapses, the carriers tend to relax to the bottom of the DOS and approach $<\varepsilon_\infty>$. At low COR, $<\varepsilon_\infty>$ lies deep in the bottom of DOS. Upon increasing COR, $<\varepsilon_\infty>$ increases towards 5.1eV, the mean energy of DOS. It is clear from the Figure 8 that in the process of carrier relaxation two groups of carriers are generated, first one whose mean energy is ~5.1eV (named as *group-I*) and the second one whose mean energy is lower than 5.1eV (named as *group-II*). This is the consequence of the slow relaxation of carriers generated inside the ordered regions of the IHL. Carriers generated inside the ordered region relax slowly and generates the set of carriers present in *group-I*. The jump rates of carriers situated around 5.1eV are higher compared to those relaxed more towards the bottom of DOS[4], i.e, jump rate of carriers in *group-I* is higher than that of the carriers in *group-II*. Thus, the generation of two groups of carriers in the process of carrier relaxation provides an additional mechanism *(in addition to the thermal and non-thermal field assisted diffusion)* for carrier diffusion. This gives an additional/increased spreading of the carrier packet or a higher rate of increase in diffusion coefficient immediately after the valley point. As the time elapses, the *group-I* carriers also gradually relax to the bottom of DOS and are merged with *group-II* carriers. Thus, this additional spreading mechanism gradually subsides once *group –I* carriers relax towards bottom of DOS. Hence, the spreading of the packet become close to that of the host lattice. This explains why at low COR (for example 30% COR, as shown in Figure 8) the carrier diffusion first increases with time and then shows an intermediate regime before it settles down to a rate of increase same as that of the host lattice. As COR increases the number of carriers in the *group-I* also increases. Therefore, the additional mechanism that results in the increased spreading of the carrier packet becomes more effective. This explains why the rate of increase in diffusion coefficient increases with the increase in COR. In such case, the carriers in *group–I* take long time to relax and merge with the carriers in the *group-II*. For example, for 60% COR the carriers in the *group-I* is distinctly visible even for the longer times as shown in the Figure 8. Upon further relaxation the number of carriers in *group-I* decreases gradually and hence the influence of additional spreading mechanism also decreases gradually but does not get eliminated as in the case of low COR. This results only in the decrease in the rate of increase of diffusion coefficient at longer times as shown in Figure 2. This explains why the intermediate regime (where the diffusion coefficient shows a plateau like region) gradually vanishes with increase in COR. It can be clearly inferred that there exists an optimum COR for which the additional mechanism that results in the increased spreading of the carrier packet becomes most effective. In our study, the optimum COR is ~85% (refer Figure 2). At this optimum concentration, the rate of increase in the diffusion coefficient immediately after the valley point is maximum (*origin of optimum concentration will be discussed later*). Increasing the COR beyond the optimum value leads to the gradual decrease in the rate of increase in the diffusion coefficient immediately after the valley point. This can be partially due to the gradual elimination of morphology dependent carrier spreading upon increasing COR. In addition, as the lattice approach towards a host lattice with low energetic disorder the contribution from the non-thermal field assisted diffusion also diminishes gradually. Even at very high COR the bottom of DOS is occupied by the highly disordered sites that act as deep site energies. These deep site energies can significantly contribute to the non-thermal field assisted diffusion. This is the reason why even at higher concentration of ordered region (~99%) the temporal evolution of diffusion coefficient is very much different compared to the pure host lattice with σ=15meV (100% COR). As COR increases these sites gradually vanishes and the temporal evolution of diffusion coefficient gradually settles towards the pure host lattice with σ=15meV (*see supporting information*).

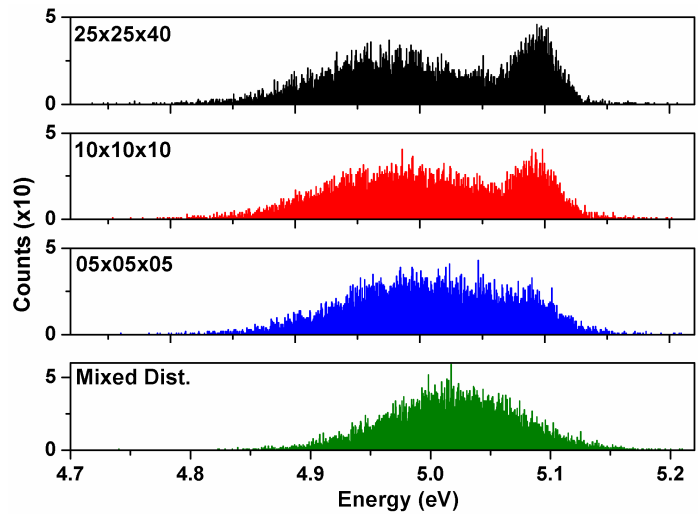

Figure 9. Histogram representing the ODS of IHL, captured at Log($t/t_0$)=1.7, upon decreasing the size of ordered regions. All simulation were carried out at E= $6.4 \times 10^5$V/cm, T=300K and COR =50%.

The above observations and explanations unambiguously establishes that the slower relaxation of the carriers generated in ordered regions that results in the formation of two groups of carriers with wide difference in jump rates is responsible for the morphology dependent diffusion and hence, the observed intriguing features of temporal evolution of diffusion coefficient in IHL. Thus, the disappearance of intriguing features of temporal evolution of diffusion coefficient in IHL upon decreasing the size of ordered region for a fixed COR (shown in figure 3) can be attributed to the concomitant disappearance of the difference in energy relaxation between the carriers generated in the ordered and highly disordered regions (as shown in figure 6). This is further justified by the decrease in the number of carriers in *group-I* upon decreasing the size of ordered region, which is clearly shown in figure 9 that shows the ODS, captured at Log($t/t_0$)=1.7, upon reducing the size of ordered region. In the mixed distribution case *group-I* carriers is not generated at all. It is also interesting to know how the steady state diffusion changes with the COR. Preliminary investigation suggests that steady state diffusivity reaches a maximum and then decreases upon increasing the COR. Increase in steady state diffusion upon increasing COR may be attributed to morphology dependent carrier spreading that enhances with increase in the COR and the contribution of host lattice sites occupied at the bottom of DOS to the non-thermal field assisted diffusion.

Simulations have been performed to understand the dependence of morphology dependent carrier spreading mechanism on applied field strengths. Figure 10(a) and figure 10(b) show the temporal evolution of diffusion coefficient simulated for two different lower electric field strengths. The features of temporal evolution of diffusion coefficient at lower electric field strengths are similar to that observed at E=6.4x$10^5$V/cm. ODS of carriers look similar to that shown before. As explained above, *group-I* and *group-II* carrier packets of different jump rates are generated upon carrier relaxation and this enhances the carrier diffusion. At low COR, the number of carriers generated in the ordered region is less and hence the carriers in the *group-I* merge faster with *group-II* resulting in the vanishing of morphology dependent carrier diffusion. Hence, at low concentration the diffusion coefficient increases at short time but decreases in the intermediate field regime and follows the temporal evolution of host lattice at long times. At higher COR, the number of carriers in *group-I* increases which enhances the strength of morphology dependent carrier diffusion. Hence, as explained above, the diffusion increases at higher rate (immediately after the valley point) and the intermediate regime vanishes at higher concentration. Similarly, the diffusion coefficient decreases with increase in COR beyond the optimum concentration of ordered region. Compared to the optimum COR (~85%) for E=6.4x$10^5$V/cm, the optimum COR for E=2x$10^5$V/cm and E=4x$10^4$V/cm are 96% and 98% respectively. The dependence of optimum COR on applied field strength suggests that the optimum COR is determined by a complex combination of many factors such as the amount of carrier generation in the ordered region, relaxation of carriers, the equilibrium energy, strength of field assisted diffusion. The amount of carrier generated in the ordered region, which depends on COR, decides the number carriers in *group-I*. The number of carriers in group-*I* should be optimum so as to have the maximum strength for the morphology dependent carrier diffusion. In addition to the number of carriers in group-*I*, the speed of relaxation of carriers in *group-II* may also influence the optimum COR. If the *group-II* carriers relax more towards the bottom of DOS, this will result in wide difference in jump rates for carriers in two groups. This will affect the strength of morphology dependent diffusion and probably the optimum COR.

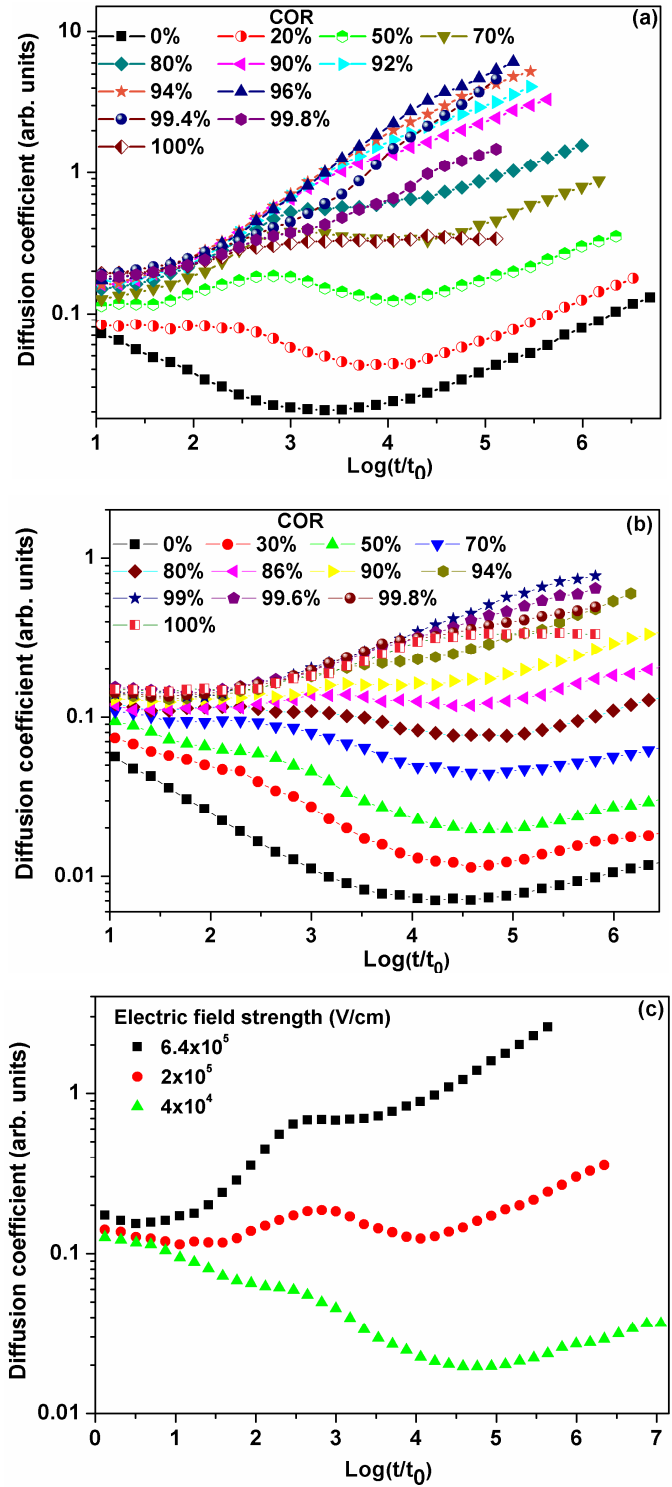

Figure 10(a) Temporal evolution of diffusion coefficient parametric with COR for IHL for (a) E=2x10$^5$V/cm (b)E=4x10$^5$V/cm. (c) Comparison of temporal evoultuion of diffuion coefficeint for various electic fields at COR=50%. Temperature used is 300K.

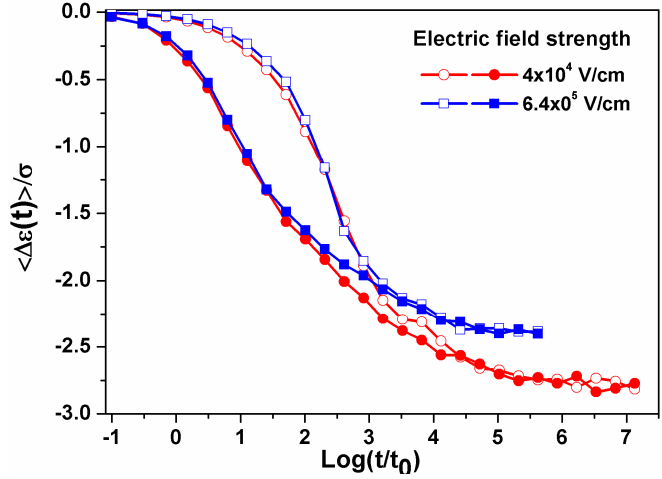

Figure 11. Comparison of temporal evolution of mean energy of carriers, at two different electric field strengths, generated specifically inside (■,●) highly disordered and (□,○) ordered regions of IHL. Maximum size of ordered region is 25x25x40 along x, y and z directions. Simulation were carried out at E= 6.4x10$^5$V/cm and T=300K for COR of 50%.

As shown in Figure 10(c), for a constant COR the enhancement of diffusion coefficient due to morphology dependent carrier spreading is higher for higher electric field strengths. Figure 11 shows the relaxation of energy of carriers generated inside ordered regions and high disorder region for lower electric field strengths. Compared to the relaxation of the energy of carriers for E=6.4x10$^5$V/cm, the initial relaxation of energy of carriers (observed for Log($t/t_0$) <2, which is responsible for the formation of two groups of carriers as explained above) observed for lower electric field strengths is almost similar. This suggests that the formation of the carrier packets with different jump rates happen similarly at lower electric field strengths under study, which is also supported by the histogram of ODS obtained for lower electric field strengths. Upon decreasing the electric field strengths, the strength of field assisted diffusion decreases and the field assisted diffusion become active at longer time. Therefore, at higher electric field strengths the field assisted diffusion is not only stronger but also act simultaneously on the carriers along with the morphology dependent diffusion. This results in the remarkable enhancement of diffusion coefficient. At lower electric field strengths, by the time the field assisted diffusion become significant the influence of morphology dependent carrier diffusion diminishes and hence the enhancement of diffusion coefficient is minimal. This is clearly observed in temporal evolution of diffusion coefficient at low COR (30%, 50%) simulated for $E=4\times10^4$ V/cm (Figure 10b). Diffusion tend to increase around $Log(t/t_0)=2$ and then proceed to attain the valley point around $Log(t/t_0)=4$, beyond which diffusion coefficient increases due to field assisted diffusion. Only at very higher COR the morphology dependent diffusion becomes strong enough to have the remarkable enhancement of diffusion coefficient. This also suggests that the field assisted diffusion may also have role in deciding the optimum COR.

The above results clearly establish a film morphology dependent mechanism of carrier diffusion that can occur in polycrystalline samples. Experimental results have shown that average size of aggregates/crystalline regions in organic semiconducting thin films is typically few tens of nanometers[27]. In this study, embedding the ordered regions of randomly varying size in a highly disordered lattice mimics a lattice with nanoscale morphology and hence, the observation of the manuscript can be relevant for the practical cases. This morphology dependent carrier diffusion can have remarkable influence on the experimental results on inhomogeneous active medium, i.e morphologically tailored active layers where the charge transport occurs through a mixture of ordered and disordered regions. Various processes like charge injection, charge transport, exciton diffusion, charge recombination etc are highly dependent on the carrier diffusion. Hence, the process of carrier diffusion has a crucial role in deciding the device performance[1]. Since most of the organic optoelectronic devices adopt morphologically tailored active layers, the proposed morphology dependent diffusion in this study is highly relevant for understanding the various optoelectronic processes and hence important for optimizing the device performance. Thus, this study highlights the need to investigate, both theoretically and experimentally, the influence of morphology dependent diffusion on the various optoelectronic properties. This may help to exploit the film morphology in a better way for designing efficient devices. Furthermore, this work makes us think that various new classes of materials such as organic–inorganic hybrids, organic-metal nano-composites etc. may involve new mechanisms of carrier diffusion and hence require further attention. In concise, this study through exemplifying the morphology dependent carrier diffusion in disordered organic materials highlights the importance of considering the influence of film morphology while interpreting the experimental results and optimizing the design of devices.

**Conclusions**

Monte Carlo simulation studies are performed to understand the influence of morphology on the diffusion of carriers. For HL, the rate of increase in the diffusion coefficient with time decreases with decrease in energetic disorder. Compared to HL, the temporal evolution of diffusion coefficient in IHL shows unexpected behavior upon decreasing the overall energetic disorder seen by the carrier. The counter intuitive temporal evolution of diffusion coefficient in IHL is explained on the basis of morphology dependent carrier spreading mechanism. The morphology dependent carrier spreading mechanism acts on carrier packet in addition to the thermal and non-thermal field assisted diffusion. This film morphology dependent carrier spreading mechanism arises due to generation of packets of carriers with wide difference in jump rates due to the slow relaxation of carriers generated in the ordered region. Such mechanism will be present in any polycrystalline system where the charge transport occurs through regions of ordered and disordered regions. Possible implications of the observed morphology dependent carrier diffusions on the operation of organic optoelectronic devices is discussed. Thus, this study not only exemplifies the influence of film morphology on carrier diffusion but also highlights the importance of the same in understanding the physics of organic photonic devices and in optimizing their performance.

**Supporting Information Available**

S1) Temporal evolution of diffusion coefficient for various concentrations of ordered regions between 99% and 100% at (a) $E=6.4 \times 10^5$ V/cm (b) $E=2 \times 10^5$ V/cm. Temperature used is 300K.

This information is available free of charge via the internet at http://pubs.acs.org

## Table of Contents – Graphic

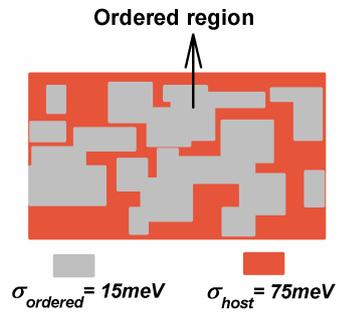 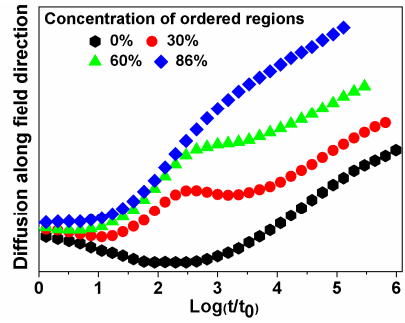